\documentstyle[psfig,subfigure]{mn}

\newif\ifAMStwofonts

\newcommand{\HI}{\mbox{H\,{\sc i}}}
\newcommand{\kms}{\mbox{km~s$^{-1}$}}
\newcommand{\am}[2]{$#1'\,\hspace{-1.7mm}.\hspace{.0mm}#2$}

\ifoldfss
  \ifCUPmtlplainloaded \else
    \NewTextAlphabet{textbfit} {cmbxti10} {}
    \NewTextAlphabet{textbfss} {cmssbx10} {}
    \NewMathAlphabet{mathbfit} {cmbxti10} {} 
    \NewMathAlphabet{mathbfss} {cmssbx10} {} 
  \fi
  \ifAMStwofonts
    \ifCUPmtlplainloaded \else
      \NewSymbolFont{upmath} {eurm10}
      \NewSymbolFont{AMSa} {msam10}
      \NewMathSymbol{\upi}     {0}{upmath}{19}
      \NewMathSymbol{\umu}     {0}{upmath}{16}
      \NewMathSymbol{\upartial}{0}{upmath}{40}
      \NewMathSymbol{\leqslant}{3}{AMSa}{36}
      \NewMathSymbol{\geqslant}{3}{AMSa}{3E}

    \fi
  \fi
\fi 

\ifnfssone
  \newmathalphabet{\mathit}
  \addtoversion{normal}{\mathit}{cmr}{m}{it}
  \addtoversion{bold}{\mathit}{cmr}{bx}{it}
  \newmathalphabet{\mathbfit} 
  \addtoversion{normal}{\mathbfit}{cmr}{bx}{it}
  \addtoversion{bold}{\mathbfit}{cmr}{bx}{it}
  \newmathalphabet{\mathbfss} 
  \addtoversion{normal}{\mathbfss}{cmss}{bx}{n}
  \addtoversion{bold}{\mathbfss}{cmss}{bx}{n}
  \ifAMStwofonts
    \ifCUPmtlplainloaded \else
      \UseAMStwoboldmath
      \makeatletter
      \new@mathgroup\upmath@group
      \define@mathgroup\mv@normal\upmath@group{eur}{m}{n}
      \define@mathgroup\mv@bold\upmath@group{eur}{b}{n}
      \edef\UPM{\hexnumber\upmath@group}
      \new@mathgroup\amsa@group
      \define@mathgroup\mv@normal\amsa@group{msa}{m}{n}
      \define@mathgroup\mv@bold\amsa@group{msa}{m}{n}
      \edef\AMSa{\hexnumber\amsa@group}
      \makeatother
      \mathchardef\upi="0\UPM19
      \mathchardef\umu="0\UPM16
      \mathchardef\upartial="0\UPM40
      \mathchardef\leqslant="3\AMSa36
      \mathchardef\geqslant="3\AMSa3E
    \fi
  \fi
\fi 

\ifnfsstwo
  \DeclareMathAlphabet{\mathbfit}{OT1}{cmr}{bx}{it}
  \SetMathAlphabet\mathbfit{bold}{OT1}{cmr}{bx}{it}
  \DeclareMathAlphabet{\mathbfss}{OT1}{cmss}{bx}{n}
  \SetMathAlphabet\mathbfss{bold}{OT1}{cmss}{bx}{n}
  \ifAMStwofonts
    \ifCUPmtlplainloaded \else
      \DeclareSymbolFont{UPM}{U}{eur}{m}{n}
      \SetSymbolFont{UPM}{bold}{U}{eur}{b}{n}
      \DeclareSymbolFont{AMSa}{U}{msa}{m}{n}
      \DeclareMathSymbol{\upi}{0}{UPM}{"19}
      \DeclareMathSymbol{\umu}{0}{UPM}{"16}
      \DeclareMathSymbol{\upartial}{0}{UPM}{"40}
      \DeclareMathSymbol{\leqslant}{3}{AMSa}{"36}
      \DeclareMathSymbol{\geqslant}{3}{AMSa}{"3E}
    \fi
  \fi
\fi 

\ifCUPmtlplainloaded \else
  \ifAMStwofonts \else 
    \def\upi{\pi}
    \def\umu{\mu}
    \def\upartial{\partial}
  \fi
\fi

\title{A multi-beam HI survey of the Virgo Cluster - two isolated \HI\ clouds ?}

\author[J. Davies et al.]
       {J. Davies$^1$, R. Minchin$^1$, S. Sabatini$^1$, W. van Driel$^2$, M. Baes$^{1,4}$, \newauthor
        P. Boyce$^1$, W. J. G. de Blok$^1$, M. Disney$^1$, Rh. Evans$^1$, V. Kilborn$^3$, \newauthor
        R. Lang$^1$,  S. Linder$^1$,
        S. Roberts$^1$, R. Smith$^1$ \\
        $^1$ University of Cardiff, School of Physics and Astronomy, 
        P.O.Box 913, Cardiff CF24 3YB, UK \\
        $^2$ Observatoire de Paris, GEPI, CNRS UMR 8111 and Universit\'e Paris 7, 
              5 place Jules Janssen, F-92195 Meudon Cedex, France\\ 
        $3$ University of Manchester, Jodrell Bank \\
        $4$ Sterrenkundig Observatorium, Universiteit Gent, Krijgslaan 281 S9, B-9000 Gent, Belgium }
\date{Draft version: 3 June 2003}

\pagerange{\pageref{firstpage}--\pageref{lastpage}}
\pubyear{2002}

\begin{document}

\maketitle

\label{firstpage}

\begin{abstract}
We have carried out a fully sampled large area ($4^{\circ} \times 8^{\circ}$) 21cm \HI\ line survey of part of the Virgo cluster using the Jodrell Bank multi-beam instrument. The survey has a sensitivity some 3 times better than the standard HIJASS and HIPASS surveys. We detect 31 galaxies, 27 of which are well known cluster members. The four new detections have been confirmed in the HIPASS data and by follow up Jodrell Bank pointed observations. One object lies behind M86, but the other 3 have no obvious optical counter parts upon inspection of the digital sky survey fields. These 3 objects were mapped at Arecibo with a smaller \am{3}{6} HPBW and a 4 times better sensitivity than the Jodrell Bank data, which allow an improved determination of the dimensions and location of two of the objects, but surprisingly failed to detect the third. The two objects are resolved by the Arecibo beam giving them a size far larger than any optical images in the nearby field. To our mass limit of $5 \times 10^{7}$  $\frac{\Delta v}{50 km s^{-1}}$ $M_{\odot}$ and column density limit of $3 \times 10^{18}$  $\frac{\Delta v}{50 km s^{-1}}$ atoms cm$^{-2}$ these new detections represent only about 2\% of the cluster atomic hydrogen mass. Our observations indicate that the \HI\ mass function of the cluster turns down at the low mass end making it very different to the field galaxy \HI\ mass function. This is quite different to the Virgo cluster
optical luminosity function which is much steeper than that in the general field. Many of the sample galaxies are relatively gas poor compared to \HI\ selected samples of field galaxies, confirming the 'anaemic spirals' view of Virgo cluster late type galaxies. The velocity distribution of the \HI\ detected galaxies is also very different to that of the cluster as a whole. There are relatively more high velocity galaxies in the \HI\ sample, suggesting that they form part of a currently infalling population. The \HI\ sample with optical identifications has a  minimum \HI\ column density cut-off more  than an order of magnitude above that expected from the sensitivity of the survey. This observed column density is above the normally expected level for star formation to occur. The two detections with no optical counterparts have very much lower column densities than that of the rest of the sample, below the star formation threshold.

\end{abstract}

\renewcommand{\baselinestretch}{2.0}
\section[]{Introduction}

With the advent of multi-beam recievers on large single dish radio telescopes it has become possible to make fully sampled 21cm surveys of large areas of sky.  The Parkes Telescope has been used to produce the HIPASS (HI Parkes All Sky Survey) of the southern sky (Staveley-Smith et al. 1996) and the Jodrell Bank Telescope is currently carrying out the HIJASS (HI Jodrell All Sky Survey) of the northern sky (Lang et al. 2002). The work we describe here is part of the HIJASS survey.

These surveys have enabled astronomers to construct, for the first time, large atomic hydrogen selected samples of galaxies (Kilborn 2000; Zwaan et al. 2003). Important results from these surveys include the derivation of the \HI\ mass function using a \HI\ selected sample (Zwaan et al. 2003), identification of High Velocity Clouds (Putman et al. 2002), the measurement of tidal streams (Boyce et al. 2001), the identification of optically faint gas rich objects (Minchin et al. 2003a) and a measurement of the total \HI\ content of clusters (Barnes et al. 1997, Waugh et al. 2002). It is a continuation of this latter
 work that we discuss in this paper. We use higher sensitivity data than Barnes et al. and Waugh et al. to study the atomic hydrogen content of the Virgo cluster (they studied the Fornax, Centaurus and Eridanus clusters).  

The Virgo cluster is by far the largest nearby grouping of galaxies. The high galaxy number density and the brightness of its prominent galaxies has attracted astronomers for centuries. It is a logical place to explore with a new instrument or at a new wavelength because the galaxies are bright and they are relatively easily resolved. The cluster environment, though, is not typical. The majority of galaxies in the Universe reside, not in large clusters but, in rather smaller groupings like the Local Group. Primary observational differences between the Virgo cluster and other environments are:
\begin{enumerate}
\item The presence of elliptical galaxies, primarily in the central regions - morphology density relation (Dressler 1980).
\item A relatively large number of dwarf galaxies (Binggeli et al. 1985).
\item The presence of inter-galactic stars - inferred from the identification of inter-galactic planetary nebula (Feldmeier et al. 1998).
\item An inter-galactic X-ray gas (Young et al. 2002).
\item Cluster spiral galaxies relatively devoid of \HI\ - the so called `anaemic spirals'  (see van den Bergh 1991, and references therein).
\item a crossing time short compared to a Hubble time (Tully et al. 2002).
\end{enumerate}
To understand how these differences have arisen we need to compare the properties of the cluster galaxy population with that of the general field.

We have previously been primarily interested in the Low Surface Brightness (LSB) dwarf galaxy population of clusters, groups and the field. A prime motivation for this survey was to try and detect extreme LSB cluster galaxies that are easier to detect via their 21cm, rather than their optical emission. A prime motivation being the CDM picture of hierarchical structure formation (White and Rees, 1978, White and Frenk, 1991) which predicts that there should be many more small dark matter halos around individual galaxies and in galaxy clusters than have been detected (Kauffmann et al., 1993, Moore et al., 1998). Most previous attempts to detect small dark matter haloes have relied upon optical observations of the luminous stellar component (the faint end of the galaxy luminosity function). It is possible that these small dark matter halos contain baryonic material, but they have not formed stars. So they may be undetectable in the optical, but detectable at 21cm.
 Blitz et al. (1999) and Braun and Burton (1999) have previously suggested that the local high velocity \HI\ clouds (HVC) detected in some 21cm surveys may infact be this population.  This then makes up the difference between the observed luminosity function and the theoretically predicted dark matter mass function. If the HVC are at Local Group ($\approx 1$ Mpc) rather than the Galactic distances ($\approx 100$ kpc) then these clouds will have masses of order a few time $10^{7}$ $M_{\odot}$. If a similar population exists in the Virgo cluster a survey like ours should be able to detect the most massive of them.

 The most important recent optical survey of the cluster was carried out by Binggeli et al. (1985). They surveyed 100 sq deg roughly centred on the dominant central elliptical galaxy M87, cataloguing some 1000 cluster galaxies down to $m_{B} \approx 18$. There were many new detections and most of these were previously unidentified cluster dwarf galaxies. There have been numerous much smaller area surveys carried out since then, to fainter magnitudes and surface brightnesses, revealing even fainter dwarf galaxies (Impey et al. 1988; Phillipps et al. 1998; Sabatini et al. 2003). We have recently compared Virgo to other less dense galaxy environments and it is clear that Virgo has relatively many more dwarf galaxies than the field (Roberts et al. 2003). What is not clear is why this is so and how it fits in with the CDM model. Is it 'nature' - the cluster has always had more dwarf galaxies (it formed that way) or 'nurture' - the dwarfs have subsequently been created in the cluster. Possible scenarios for the latter are galaxy harassment (Moore et al., 1998) which also explains the intra-cluster material (stars and gas). The relative lack of \HI\ in Virgo cluster spirals is thought to be due to ram pressure stripping by the intra-cluster gas.

By observing the cluster at 21cm we hope to determine what the most important evolutionary precesses are that act on cluster, but not field galaxies and visa-versa.
With a multi-beam \HI\ survey we can now make direct comparisions between the \HI\ properties of field and cluster galaxies selected by their \HI\ rather than their optical characteristics. In the following sections we describe the Jodrell Bank multi-beam survey of the Virgo cluster, object detection, follow up Arecibo observations and the characteristics of the detected objects.

\section[]{The HIJASS Data}
The HIJASS uses a 4 beam 21cm receiver mounted on the 76m Lovell telescope (beam FWHM $\approx$ 12 arc min). A 64 MHz bandpass with 1024 channels is used which gives a velocity range of about -1000 to 10,000 km s$^{-1}$. Interference removes part of the band width between about 5000 and 7000 km s$^{-1}$. The Virgo cluster has a recession velocity of about 1050 km s$^{-1}$ and a velocity dispersion of about 700 km s$^{-1}$ (Binggeli et al. 1993) and so we have restricted our object identifications to those objects with velocities between 500 and 2500  km s$^{-1}$. The lower limit avoids possible confusion with high velocity clouds. The upper limit includes galaxies thought to be falling into Virgo from the far side (Binggeli et al. 1993). In the calculation of galaxy masses we have used a distance of 16 Mpc for all of the galaxies. This corresponds to the cepheid distance modulus (31.04) calculated for M100 using HST data (Graham et al. 1996). Given the range of galaxy velocities and the physical size of the cluster this could lead to distance errors of order a factor of two, or a factor of four in calculated \HI\ mass.

The Virgo survey data were assembled by scanning the receivers in a  declination strip of just over $8^{o}$ along $4^{o}$ in RA, each strip being separated by 10 arc min. The integration time per point is nominally 3500 sec, some 9 times longer than that for the standard survey product. The actual fully sampled scan area extends from  RA(J2000)$\approx12.25hr-12.5$hr and dec(J2000)$\approx12^{o}-20^{o}$. M87 (RA(2000)$\approx12.5$hr, dec(J2000)$\approx12.4^{o}$) lies in one corner of the cube and we have observed one quadrant of the cluster. Bandpass correction and calibration have been carried out using the methods described in Barnes et al. (2001). The data have then been gridded into a three-dimensional data cube ($\alpha$, $\delta$, $V_{\odot}$), with a spatial pixel size of 4 arc min. The rms noise level in each spectra is about 4 mJy beam$^{-1}$. This compares with $\approx 14$  mJy beam$^{-1}$ for the standard HIPASS data used by Barnes et al. (1997) and Waugh et al. (2002) for their nearby cluster survey. The standard HIJASS survey has a rms noise of about 13 mJy beam$^{-1}$. We thus expect to be able to detect \HI\ massess about a factor of 3-4 lower, at a given distance, than the standard survey data. The FWHM velocity resolution is $\approx 50$ km s$^{-1}$.

\begin{figure*}
\centerline{\psfig{figure=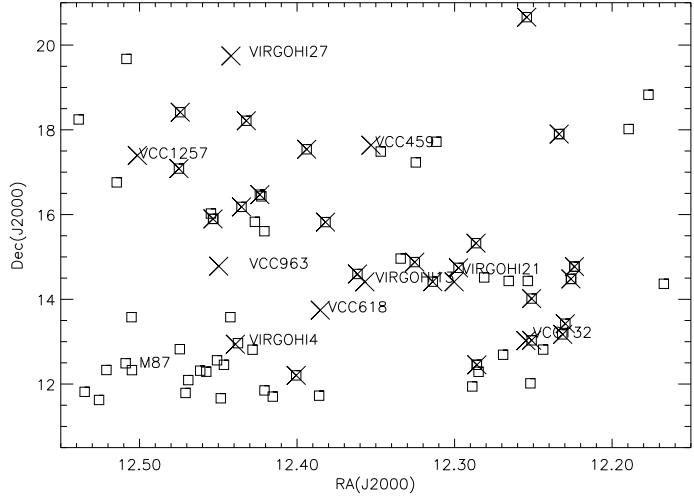,width=10cm}}
\caption
{The positions of HIJASS \HI\ detections are marked by crosses. Objects in the correct velocity range listed in the NGC, UGC or IC catalogues are marked as boxes.  \HI\ detections not in the NGC, UGC or IC catalogues have been labeled. Declination is in degrees and RA in decimal hours.}
 \end{figure*}

\section[] {Object detection in HIJASS}
We carried out two methods of object detection. Firstly we ran through the data cube looking for objects simply by eye. These appear as bright regions in the cube at different velocities, in just the same way that images are identified visually on a CCD. Searching by eye led to a list of 30 objects. Almost all of these objects were subsequently identified as relatively bright galaxies listed in LEDA (Lyon-Meudon Extragalactic DAtabase). The exception were 3 objects that we will discuss in more detail below. 

The second object detection method was an automated galaxy finder POLYFIND. This cross-correlation with templates method is described in Davies et al. (2000). The program has since been adapted to run directly on data cubes rather than extracted spectra. The program initially looks for peak values above a pre-defined value (in this case 4.5$\sigma$) and then cross-correlates with templates accepting the best fit as long as the correlation coefficient is above 0.75 and the total signal-to-noise of the detection is above 3. An extensive discussion of the automated galaxy finder and the results of running it on simulated data can be found in Minchin et al. (2003c). With a 4.5$\sigma$ peak detection a 50 km s$^{-1}$ source corresponds to a \HI\ mass of about $5 \times 10^{7}$ $M_{\odot}$ at the distance of the Virgo cluster (16 Mpc). This translates into a column density limit (a limiting mass galaxy filling the beam at the distance of Virgo) of $3 \times 10^{18}$ atoms cm$^{-2}$. Minchen et al. (2003a) show that data selected in this way produces a peak flux, rather than a total flux limited sample. So, our survey limits also depend on the velocity width of the source. Thus our survey detection limits are $5 \times 10^{7}$ $\frac{\Delta v}{50 km s^{-1}}$ $M_{\odot}$ and $3 \times 10^{18}$ $\frac{\Delta v}{50 km s^{-1}}$ atoms cm$^{-2}$ for the mass and column density respectively.

The automated technique detected 22 sources, 19 of which were relatively bright sources in the 'by eye' sample. Again we were left with 3 (different) objects that did not appear to be associated with previously known (optical) sources.
 In February 2003 we carried out follow up observations at Jodrell Bank of the six unidentified sources (three from the 'by-eye' sample and three from the POLYFIND sample). Two of these sources were rejected as noise while four were confirmed. This left a final sample of 31 objects.
A list of all the galaxies detected and their derived parameters is given in table 1.

We have compared our calibration with both the data given in LEDA (from a wide variety of sources) and that of the HIPASS survey. The relations we find are \\
\begin{center}
\[ \log{F_{{\small VIRGOHI}}}=0.9 \pm 0.1 \log{F_{{\small LEDA}}}-0.11\pm 0.11 \]

\[ \log{F_{{\small VIRGOHI}}}=1.0 \pm 0.2 \log{F_{{\small HIPASS}}}-0.04 \pm 0.19 \] 
\end{center}
where F is the flux integral (Jy km s$^{-1}$). There is about a 30\% difference between our fluxes and those given in LEDA, but the agreement is very good with the HIPASS data. Only twelve of our galaxies were detected at high enough signal to noise in the HIPASS data to make this comparison.

\begin{figure*}
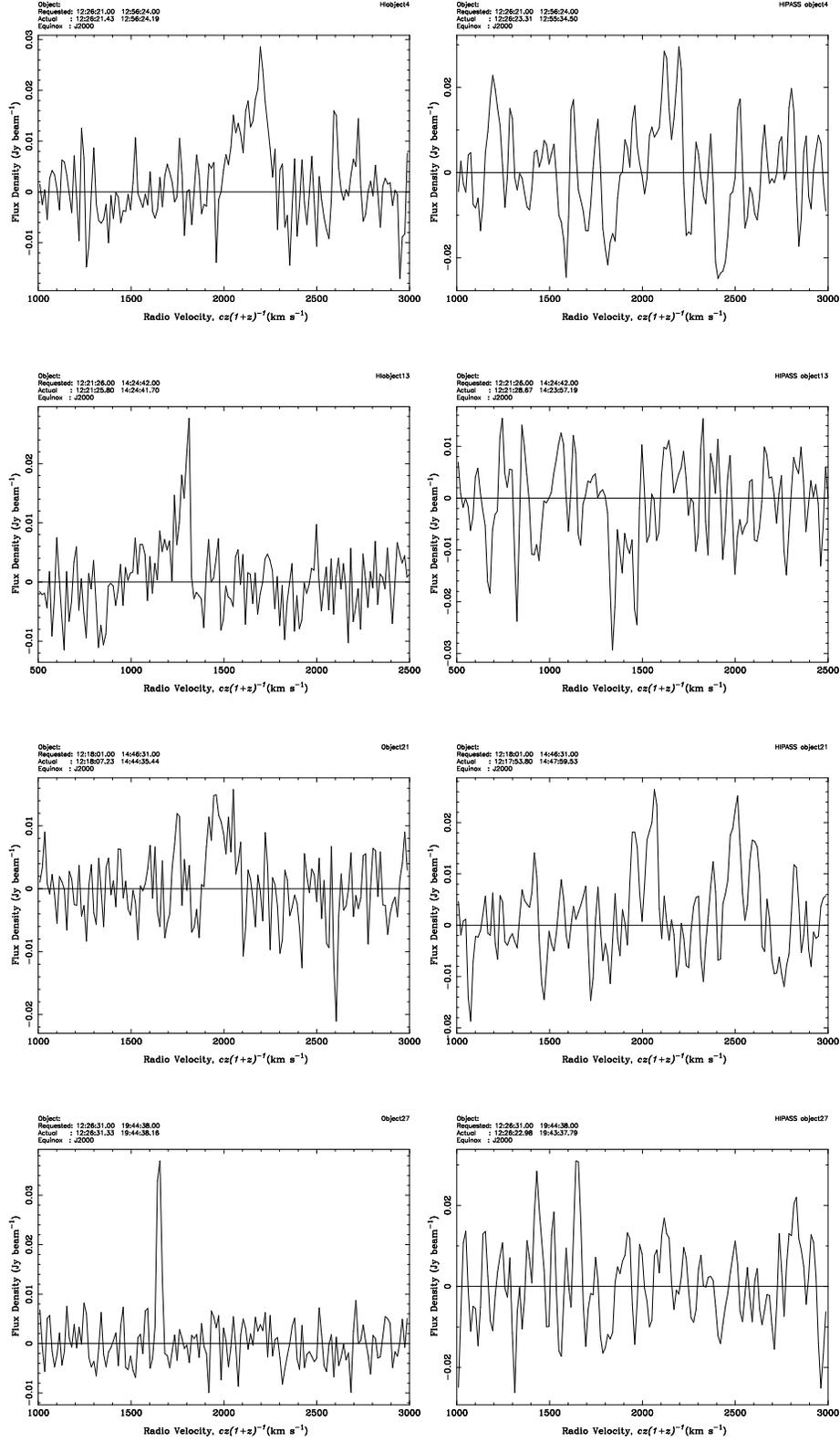

\subfigure{\psfig{figure=hiobject4.ps,width=6cm,angle=-90}}
\subfigure{\psfig{figure=hipass4.ps,width=6cm,angle=-90}}
\subfigure{\psfig{figure=hiobject13.ps,width=6cm,angle=-90}}
\subfigure{\psfig{figure=hipass13.ps,width=6cm,angle=-90}}
\subfigure{\psfig{figure=21.ps,width=6cm,angle=-90}}
\subfigure{\psfig{figure=hipass21.ps,width=6cm,angle=-90}}
\subfigure{\psfig{figure=27.ps,width=6cm,angle=-90}}
\subfigure{\psfig{figure=hipass27.ps,width=6cm,angle=-90}}
\caption
{HI spectra of objects from the HIJASS survey with no initial optical identification. On the left are the spectra from HIJASS and on the right from HIPASS. From top to bottom objects 4, 13, 21 and 27.}
 \end{figure*}

\begin{table*}
\begin{tabular}{l|c|c|c|c|c|c|c|c}
Name     & Type &  v    &  $W_{20}$ & $F_{T}$  & $\log{M_{HI}}$ & $M_{B}$ & $(M_{HI}/L_{B})$ & Col Den ($\times 10^{20}$)\\
         &  & $(km s^{-1})$ & $(km s^{-1})$ & $(Jy$ $km s^{-1})$ & $M_{\odot}$ &  &   & ($atoms$ $cm^{-2}$)  \\ \hline
M100   & SAB & 1564 & 276 & 21.5 & 9.1 & -21.2 & 0.03 & 2.6\\
M99  & SA & 2398 & 268 & 66.7 & 9.6 & -20.9 & 0.1 & 16.6\\
NGC4189 & SAB   & 2095 & 262 & 7.7 & 8.7 & -18.9 & 0.01 & 10.0\\
NGC4192A & SA  & 2042 & 150 & 6.2 & 8.6 & -15.0 & 2.6 & - \\
NGC4193  & SAB & 2474 & 441 & 16.1 & 9.0 & -18.4 & 0.2 & 24.0\\
NGC4204  & SB & 851 & 101 & 17.2 & 9.0 & -17.2 & 1.0 & 8.7\\
NGC4206  & SA & 703 & 292 & 30.4 & 9.3 & -19.4 & 0.4 & 7.2\\
NGC4237  & SAB & 905 & 178 & 2.5 & 8.2 & -18.9 & 0.03 & 4.4\\
NGC4262  & SB & 1489 & 163 & 4.3 & 8.4 & -18.7 & 0.1& 8.7\\
NGC4302  & Sc & 1142 & 372 & 22.9 & 9.2 & -19.9 & 0.1& 7.9 \\
NGC4344  & Sp/BCD & 1143 & 74 & 0.8 & 7.7 & -17.9 &  0.02 & 3.5\\
NGC4351  & SB & 2297 & 111 & 3.8 & 8.4 & -18.2 & 0.1  & 8.7\\
NGC4383  & Sa & 1700 & 228 & 38.2 & 9.4 & -18.7 & 0.6 &  75.7\\
NGC4394  & SB & 911 & 183 & 4.7 & 8.5 & -19.2 & 0.04 & 3.2\\
NGC4405  & SA & 1737 & 119 & 1.8 & 7.6 & -18.3 & 0.04 &  4.8\\
NGC4450  & SA & 1839 & 103 & 4.8 & 8.5 & -20.4 & 0.01 & 1.4\\
UGC07237  & Sm & 2257 & 141 & 3.3 & 8.3 & -14.3 & 2.7& 47.8 \\
IC3049  &  ImIII-IV & 2425 & 89 & 1.3 & 7.9 & -16.3 & 0.2 & 10.0\\
IC3061  & SBc & 2325 & 380 & 10.5 & 8.8 & -18.0 & 0.3 & 15.1\\
IC3099  & Sbc & 2117 & 227 & 3.7 & 8.4 & -17.4 & 0.2 & 8.3\\
IC3365   & Im & 2332 & 135 & 3.5 & 8.3 & -16.8 & 0.03 & 7.9\\
IC3391  & Scd & 1694 & 101 & 2.2 & 8.1 & -17.4 & 0.1 & 12.0\\
VCC0132  & SB? & 2065 & 37 & 0.6 & 7.6 & -14.8 & 0.3 &  3.3\\
VCC0459  & BCD & 2077 & 167 & 2.4 & 8.2 & -16.7 & 0.4 & 50.0\\
VCC0618  & I? & 1874 & 57 & 0.55 & 7.5 & -14.5 & 0.4 &  10.5\\
VCC0963  & I? & 1848 & 49 & 0.26 & 7.2 & -13.6 & 0.4 & 12.0\\
VCC1257  & I? & 2467 & 150 & 3.5 & 8.3 & -14.4 & 2.6 &  16.6\\
VIRGOHI4  & - & 2129 & 254 & 4.1 & 8.4 & -& - & - \\
VIRGOHI13  & - & 1274 & 100 & 2.4 & 8.2 & - & - & -\\
VIRGOHI21  & - & 1966 & 142 & 2.8 & 8.2 & -& -& - \\
VIRGOHI27  & - & 1652 & 45 & 1.3 & 7.9 & -& - &- \\
\end{tabular}
\caption{Properties of the galaxies detected in HIJASS. [1] Name, [2] Morphological type from NED, [3] Line of sight velocity, [4] Velocity width, [5] Line flux, [6] Mass of atomic hydrogen, [7] Absolute B magnitude calculated using the apparent B magnitude given in LEDA and assuming a distance of 16 Mpc, [8] Hydrogen mass to blue light ratio, [9] Mean hydrogen column density.}
\end{table*}

\section[] {Description of objects detected}
As stated above both the 'by eye' and automated detection methods predominately detected previously catalogued relatively bright galaxies. The positions of these objects are shown in Fig. 1, where the centre of the cluster (M87) is at the bottom left.

\begin{figure*}
\subfigure{\psfig{figure=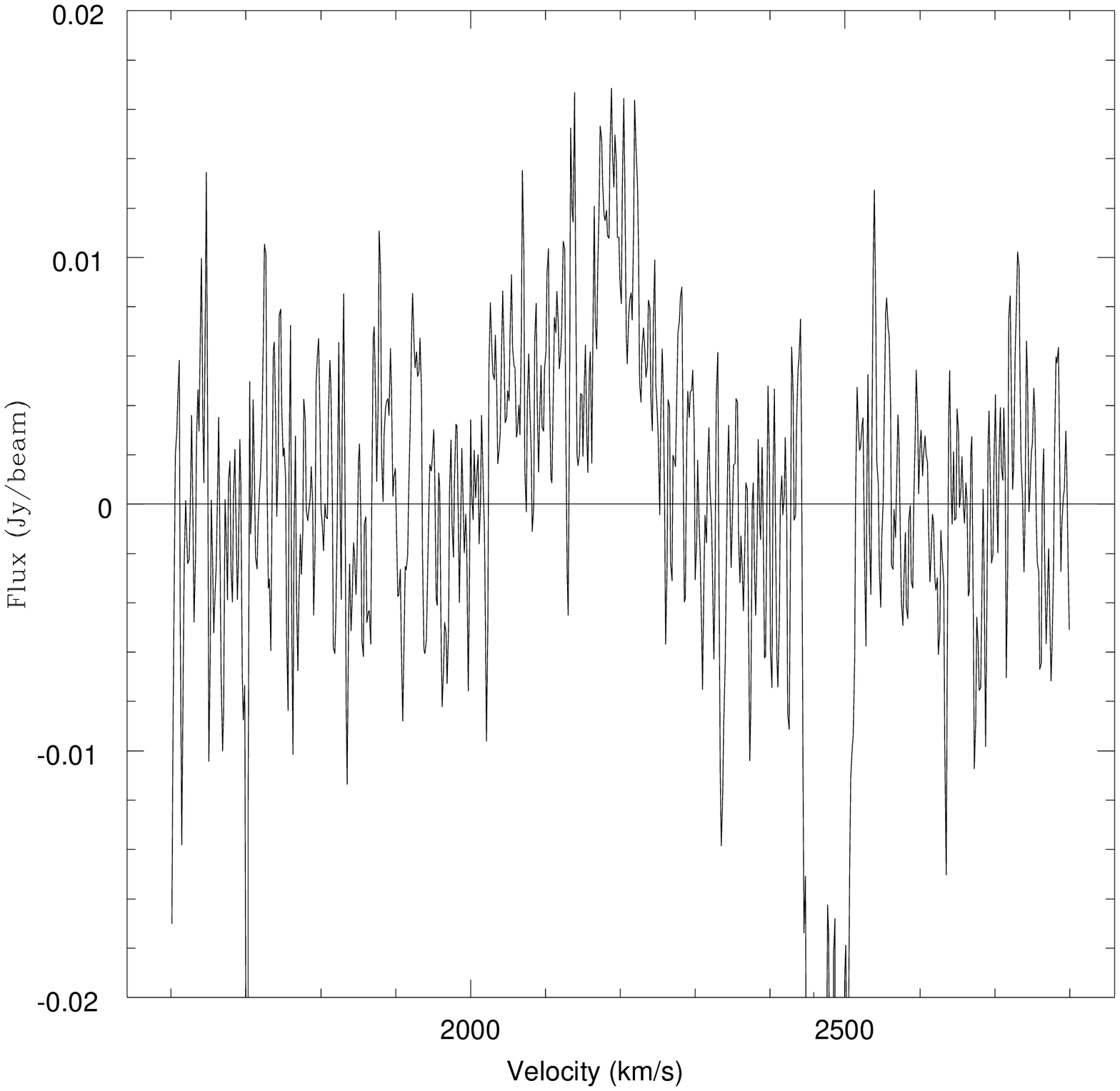,width=6cm}}
\subfigure{\psfig{figure=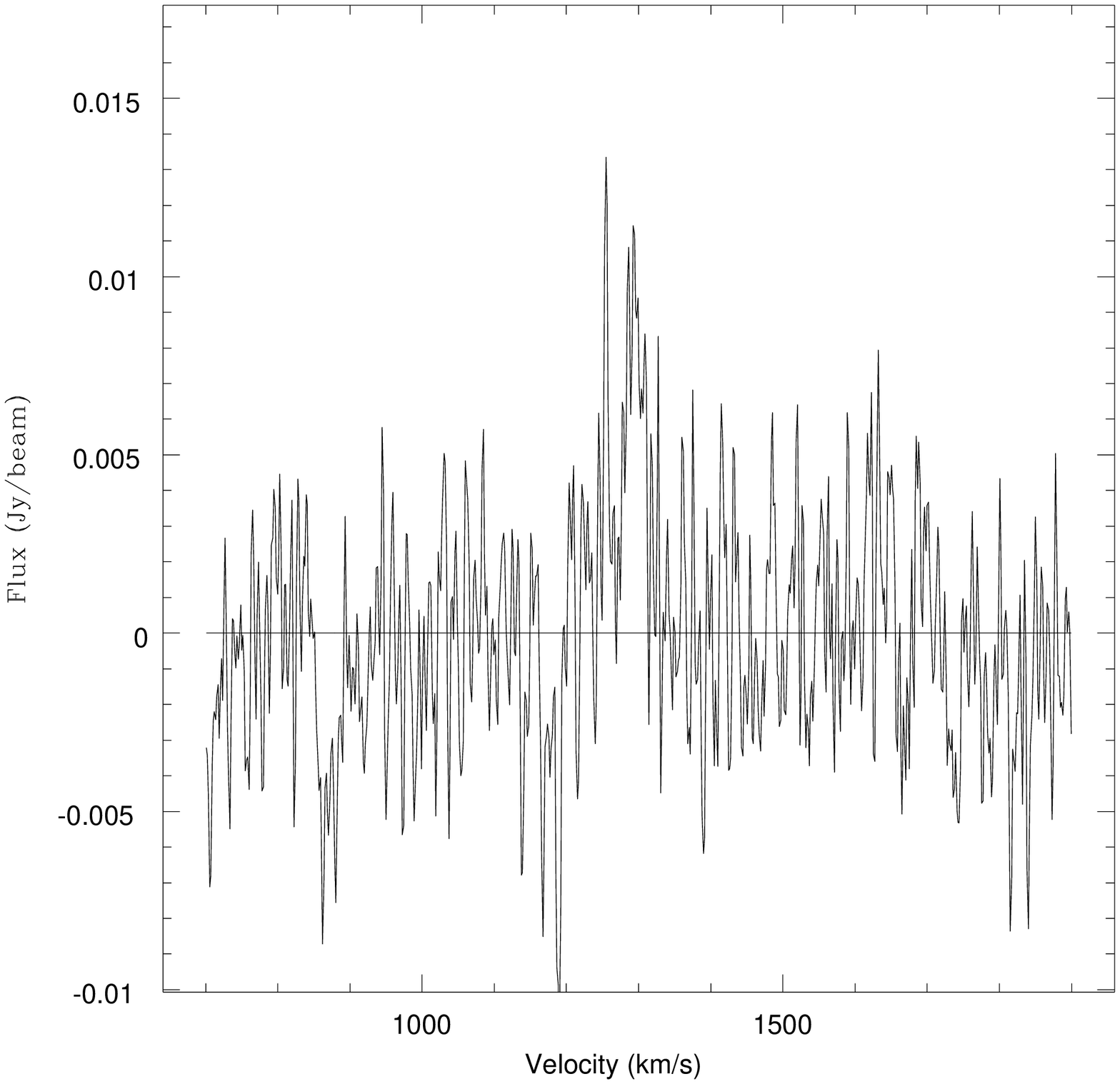,width=6cm}}
\subfigure{\psfig{figure=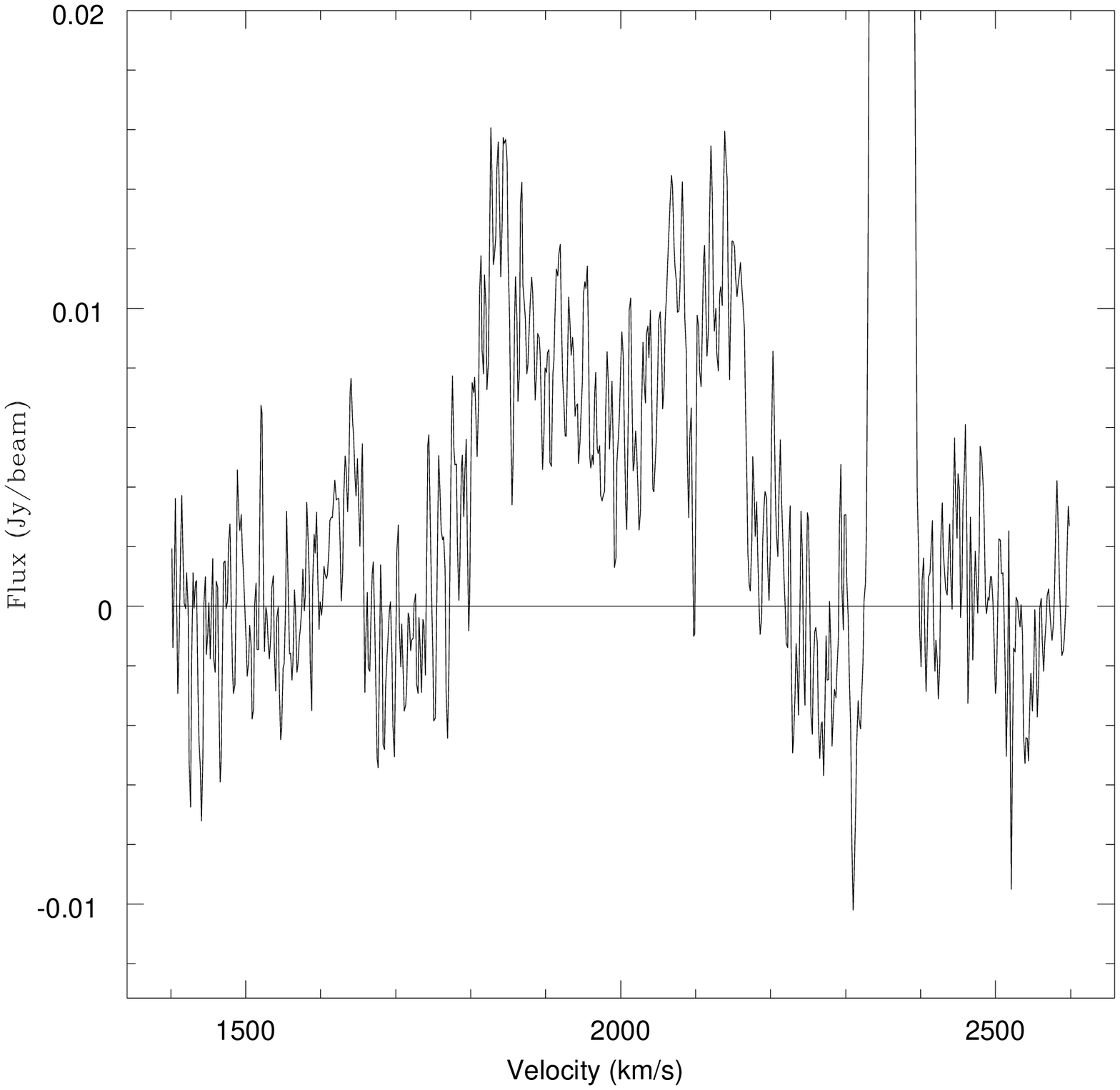,width=6cm}}
\subfigure{\psfig{figure=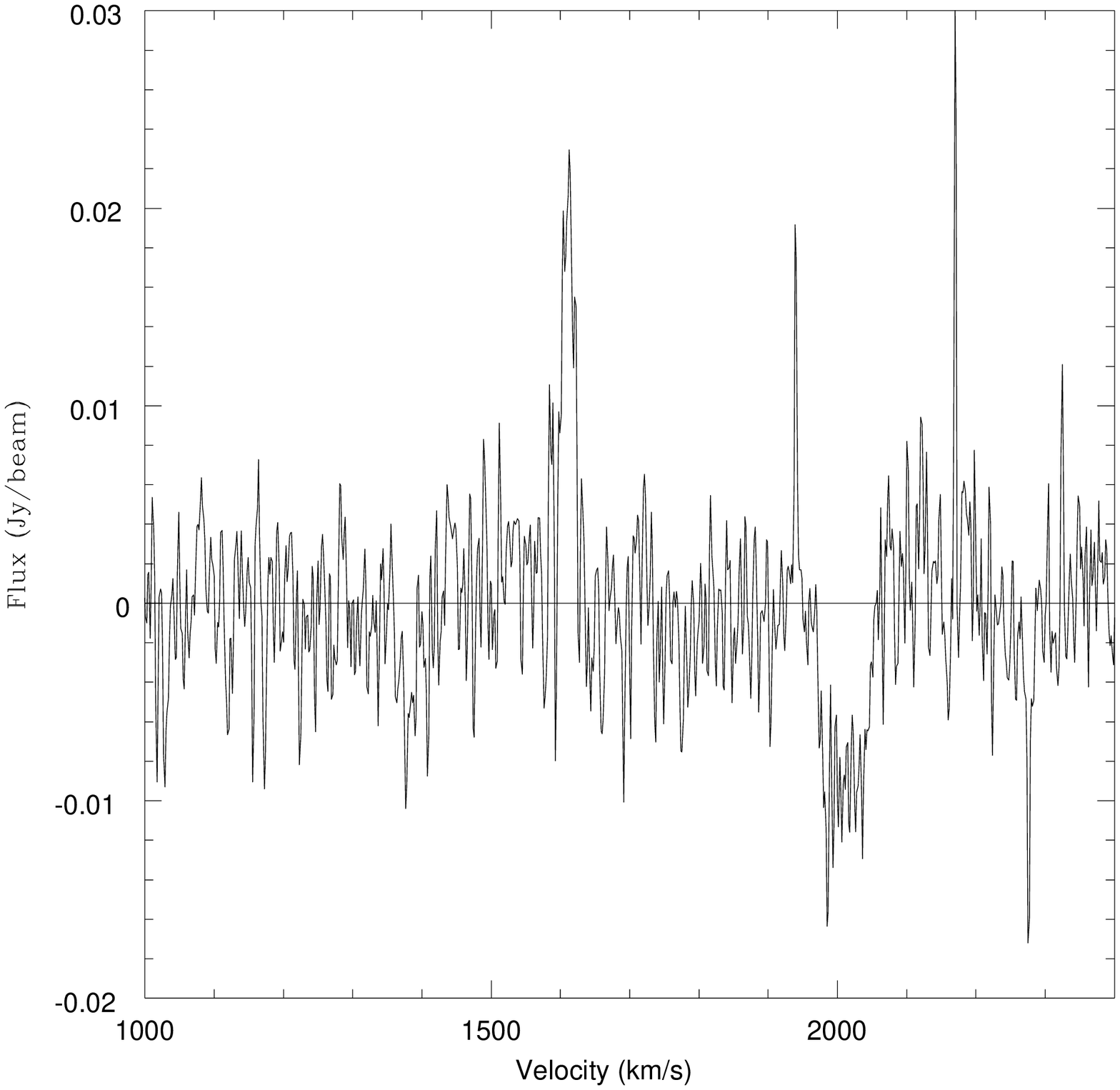,width=6cm}}
\caption
{Jodrell Bank follow up spectra of the four sources with no obvious optical identification. From left to right VIRGOHI4, VIRGOHI13, VIRGOHI21 and VIRGOHI27.}
 \end{figure*}

Although this is a 2-dimensional picture of a 3-dimensional structure it does appear that the gas rich objects avoid the cluster centre. This is another confirmation of the well known morphology density relation (gas rich late type galaxies tend to reside on the outskirts of clusters, see Dressler 1980). A similar result has been found for the Fornax cluster by Waugh et al. (2002) and for other clusters by Giovanelli \& Haynes (1985), Cayatte et al. (1994) and Bravo-Alfaro et al. (2000). \HI\ detections that correspond to optical detections (NGC, UGC and IC) are all late type galaxies. Bright galaxies with no \HI\ detection are almost all early type galaxies. There were 8 late type galaxies $(T>2)$ that potentially should have been detected in \HI. Previous \HI\ measurements (taken from NED) 
of  IC 3065, IC 3077, NGC 4371, NGC 4479 and IC 3473 indicated that they were all too faint to be detected by this survey. UGC 7170 and UGC 7186 should have been detected, but they  have velocities very close to our upper limit ($\approx$ 2400 km s$^{-1}$) and were missed. Similarly  UGC 7249 should have been detected but its velocity (622  km s$^{-1}$) is again very close to the lower limit and was not detected.

In fig. 2 we show the HIJASS \HI\ spectra for the 4 sources with no obvious optical identifications. To confirm these detections we have also looked at the pre-release HIPASS data of Virgo (right column, fig 2.). These are very marginal detections in the noisier HIPASS data, but given the benefit of knowing where to look, all objects except VIRGOHI13 are identifiable in HIPASS. The Virgo HIJASS data indicate that there are some additional sources to be found by going deeper compared to HIPASS, but future detections will not amount to a huge amount of extra \HI, unless the \HI\ mass function takes a dramatic up-turn at some point below our detection limit. In fig. 3 we show the followup spectra obtained at Jodrell Bank by integrating at the HIJASS position. These confirm the HIJASS detections. \\
\begin{figure*}
\subfigure{\psfig{figure=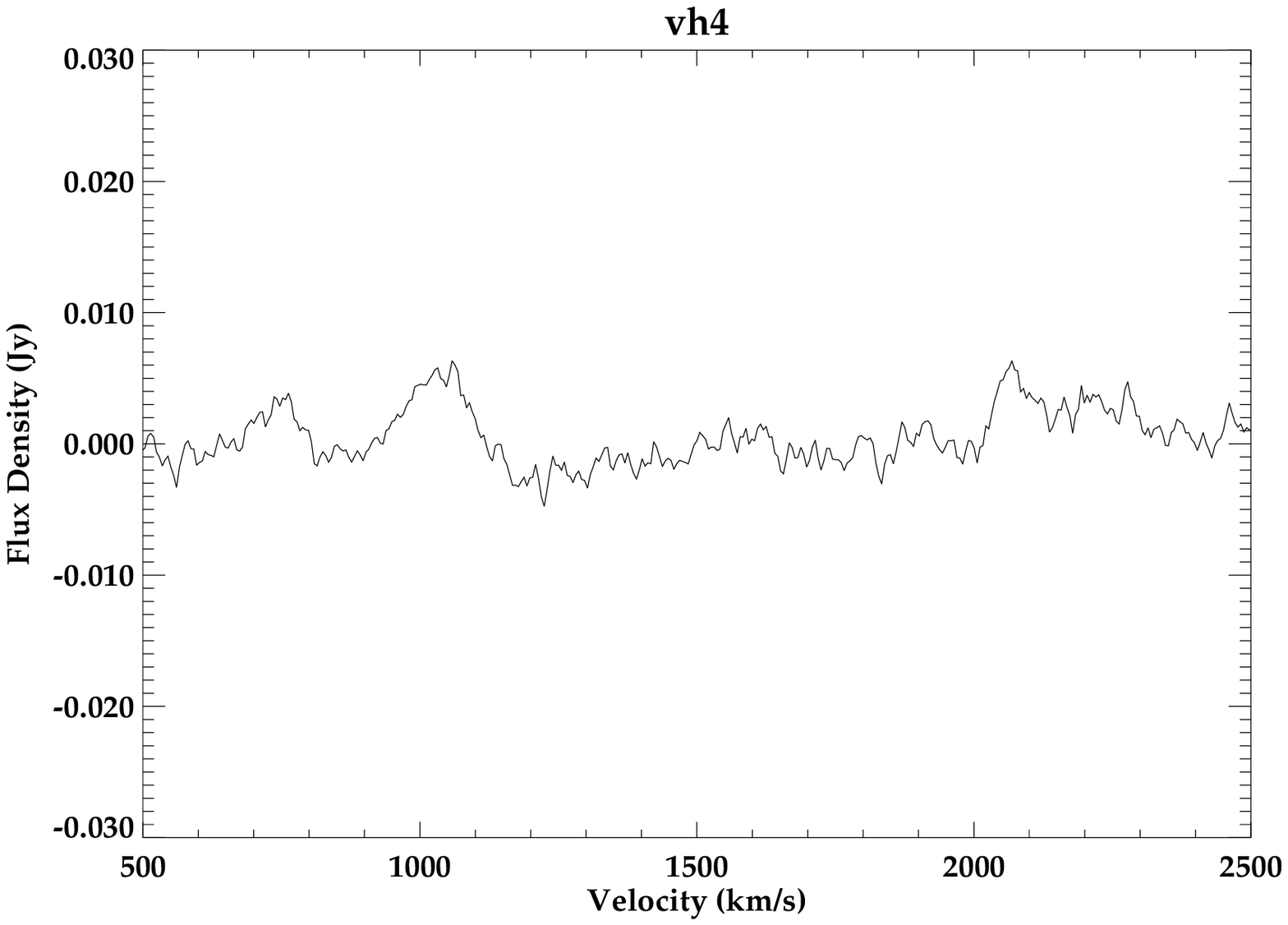,width=6cm}}
\subfigure{\psfig{figure=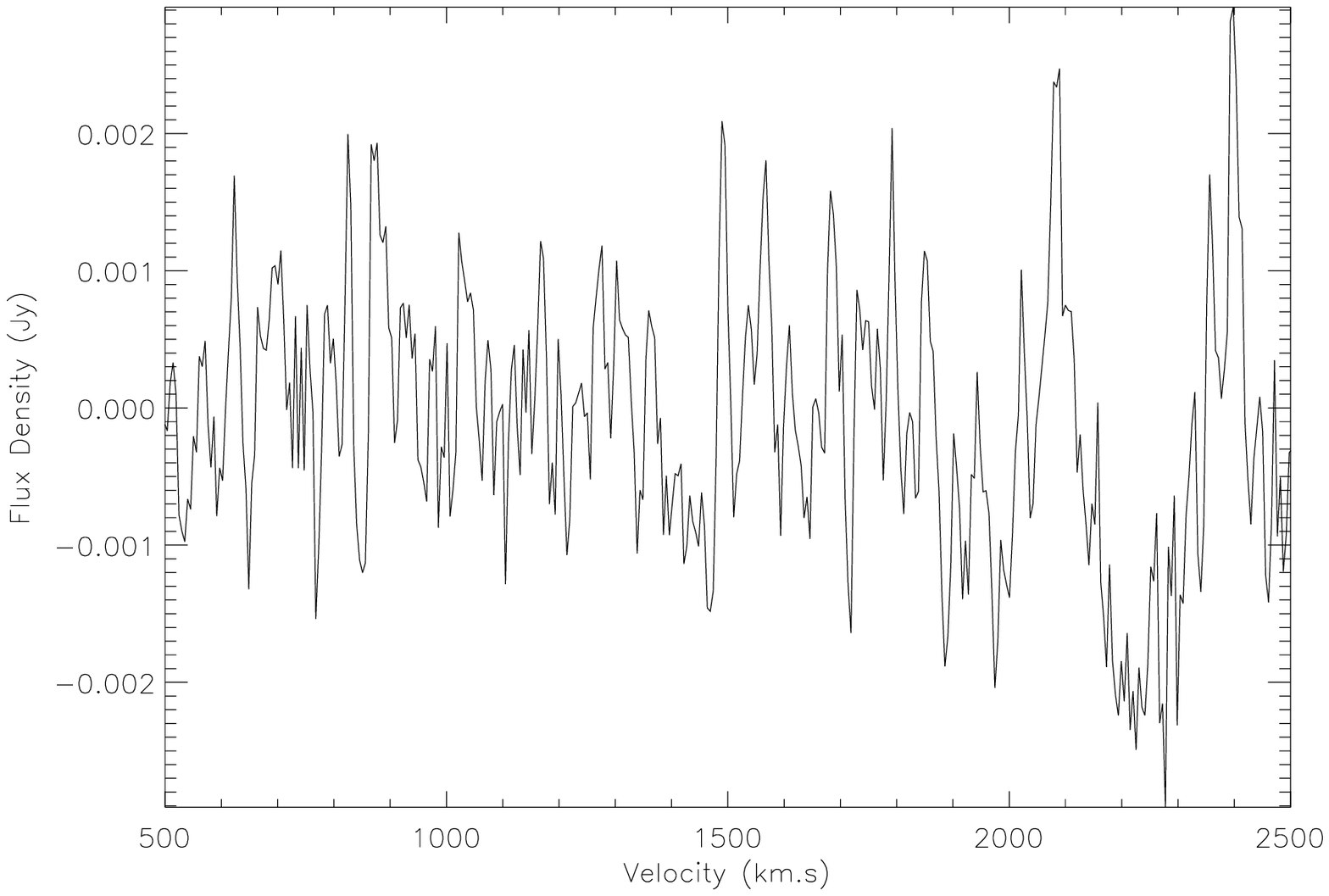,width=6cm}}
\subfigure{\psfig{figure=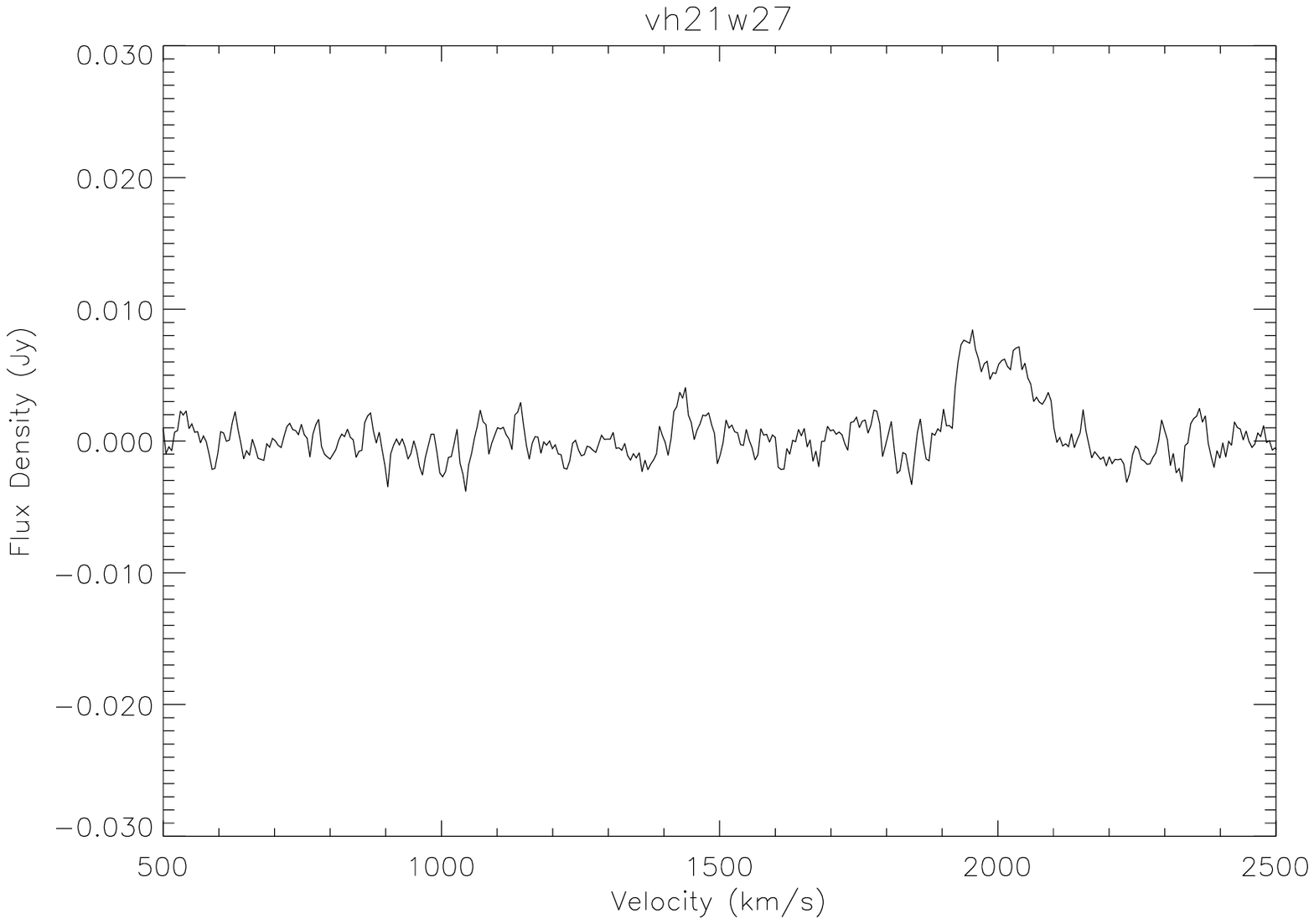,width=6cm}}
\subfigure{\psfig{figure=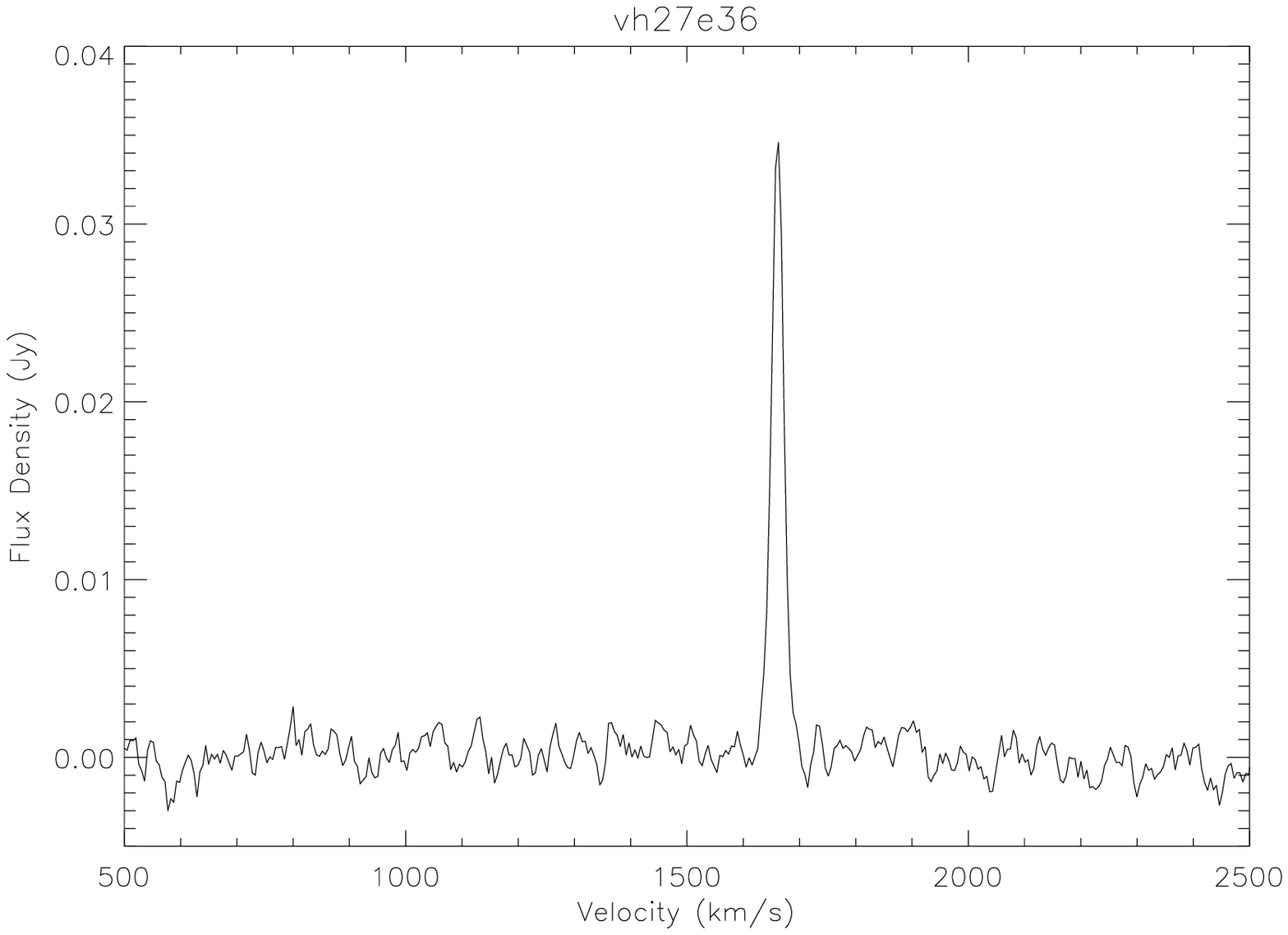,width=6cm}}
\caption
{Arecibo follow up spectra of the four sources with no obvious optical identification. From left to right VIRGOHI4, VIRGOHI13, VIRGOHI21 and VIRGOHI27. For VIRGOHI4 and VIRGOHI13 the spectra correspond to the HIJASS position for VIRGOHI21 and VIRGOHI27 the spectra are those at peak flux.}
 \end{figure*}

\subsection[] {The Arecibo Data}
The 4 \HI\ detections  without optical counterparts were re-observed using the 305-m Arecibo radio telescope. These observations have 3 times better spatial resolution (\am{3}{6} HPBW) and 4 times better sensitivity ($\sim$1 mJy rms at 15 \kms\ resolution). Each object was maped using a number of positions centred on and then off-set from the HIJASS position (see section 5).

Data were taken with the L-Band Narrow receiver using nine-level sampling with two of the 2048 lag subcorrelators set to each polarization channel. All observations were taken using the position-switching technique, with the blank sky (or OFF) observation taken for the same length of time, and over the same portion of the Arecibo dish as was used for the on-source (ON) observation. Each 3min+3min ON+OFF pair was followed by a 10s ON+OFF observation of a well calibrated noise diode. The overlaps between both sub-correlators with the same polarization allowed a wide velocity search while ensuring an adequate coverage in velocity. The velocity search range was -1000 to 10,000 \kms. The instrument's HPBW at 21 cm is \am{3}{6}$\times$\am{3}{5} and the pointing accuracy is about 15$''$.

Using standard IDL data reduction software available at Arecibo, corrections were applied for the variations in the gain and system temperature with zenith angle and azimuth, a baseline of order one to three was fitted to the data, excluding those velocity ranges with \HI\ line emission or radio frequency interference (RFI), the velocities were corrected to the heliocentric system, using the optical convention, and the polarisations were averaged.  For all spectra the rms noise level was determined and for the detected lines the central velocity, widths at, respectively, the 50\% and 20\% level of peak maximum, and the integrated flux  were determined. All data were boxcar smoothed to a velocity resolution of 15 \kms\  for analysis.

First, for each object spectra were taken at the nominal centre position of the HIJASS detection and then a search was made for \HI\ line emission around this location, at positions which are listed in table 2, until an estimate of the dimensions and positions of the sources could be made. Listed in table 2 for each pointing centre are the rms noise level, as well as the centre velocity, the $W_{50}$ line width and the integrated flux, $I_{HI}$ of the detected \HI\ lines. In each case the apertures giving a detection are surrounded by apertures that have no detection. In this way we can rule out the possibility of contamination by a bright source that might appear in a side lobe of the telescope.

\begin{table*}
\begin{tabular}{l|l|l|l|r|r|r}
Obj. & \multicolumn{2}{c}{offset} &  rms &  $I_{HI}$ &  $W_{50}$ &  $V_{HI}$ \\ 
 &  \multicolumn{2}{c}{(arcmin)} & (mJy) & (Jy km/s) & (km/s) & (km/s) \\    
\hline
VIRGOHI4  &      &      & 1.55 &  0.96  & 360 &  2200  \\
\vspace{-2mm} \\
VIRGOHI13 &      &      & 0.85 &    --  &  -- &    --  \\ 
VIRGOHI13 & N9.0 &      & 1.53 & [0.72  & 144 &  1037] \\ 
VIRGOHI13 & N5.4 & W3.6 & 0.87 &    --  &  -- &    --  \\
VIRGOHI13 & N5.4 &	   & 0.91 &  0.21  & 117 &  1269  \\ 
VIRGOHI13 & N5.4 & E3.6 & 1.09 &    --  &  -- &    --  \\
VIRGOHI13 & N2.7 & W3.6 & 0.93 &    --  &  -- &    --  \\
VIRGOHI13 & N2.7 &      & 0.94 &    --  &  -- &    --  \\
VIRGOHI13 & N2.7 & E3.6 & 1.04 &    --  &  -- &    --  \\
VIRGOHI13 &      & W5.4 & 0.94 &    --  &  -- &    --  \\
VIRGOHI13 &      & W2.7 & 0.87 &    --  &  -- &    --  \\
VIRGOHI13 &      & E2.7 & 0.89 &    --  &  -- &    --  \\ 
VIRGOHI13 &      & E5.4 & 0.91 &    --  &  -- &    --  \\  
VIRGOHI13 & S2.7 &      & 0.87 &    --  &  -- &    --  \\
VIRGOHI13 & S3.6 & W3.6 & 0.84 &    --  &  -- &    --  \\
VIRGOHI13 & S3.6 & E3.6 & 0.89 &    --  &  -- &    --  \\
VIRGOHI13 & S5.4 &      & 1.29 &    --  &  -- &    --  \\  
\vspace{-2mm} \\
VIRGOHI21 &      &      & 0.99 &  0.32  & 175 &  2001  \\ 
VIRGOHI21 &      &      & 0.84 &  0.35  & 263 &  1885  \\ 
VIRGOHI21 & N7.2 & W2.7 & 0.89 &    --  &  -- &    --  \\
VIRGOHI21 & N5.4 &      & 0.86 &  [0.20 & 88 & 2000]  \\ 
VIRGOHI21 & N3.6 & W5.4 & 0.90 &    --  &  -- &    --  \\
VIRGOHI21 & N3.6 & W2.7 & 0.91 &  0.43  & 155 &  2082  \\ 
VIRGOHI21 & N3.6 &      & 0.89 &    --  &  -- &    --  \\
VIRGOHI21 &      & W5.4 & 1.07 &    --  &  -- &    --  \\
VIRGOHI21 &      & W2.7 & 1.29 &  1.00  & 239 &  1980  \\ 
VIRGOHI21 &      & E2.7 & 1.14 &    --  &  -- &    --  \\
VIRGOHI21 &      & E5.4 & 0.90 &    --  &  -- &    --  \\
VIRGOHI21 & S3.6 & W5.4 & 1.00 &    --  &  -- &    --  \\ 
VIRGOHI21 & S3.6 & W2.7 & 0.89 &  0.59  & 165 &  1977  \\ 
VIRGOHI21 & S3.6 &      & 0.91 &  0.64  & 157 &  1947  \\ 
VIRGOHI21 & S3.6 & E2.7 & 0.84 &    --  &  -- &    --  \\
VIRGOHI21 & S5.4 &      & 0.86 &    --  &  -- &    --  \\
VIRGOHI21 & S7.2 & W2.7 & 0.88 &    --  &  -- &    --  \\
\vspace{-2mm} \\
VIRGOHI27 &      &      & 0.77 &  0.43  &  52 &  1658  \\ 
VIRGOHI27 & N7.2 & E3.6 & 0.89 & [0.22  &  93 &  1561] \\ 
VIRGOHI27 & N3.6 &      & 0.98 &    --  &  -- &    --  \\
VIRGOHI27 & N3.6 & E3.6 & 0.98 &  0.25  &  55 &  1657  \\
VIRGOHI27 & N3.6 & E7.2 & 0.91 &    --  &  -- &    --  \\
VIRGOHI27 &      & W3.6 & 0.93 &    --  &  -- &    --  \\
VIRGOHI27 &      & E3.6 & 0.97 &  0.99  &  50 &  1660  \\ 
VIRGOHI27 &      & E7.2 & 0.88 &    --  &  -- &    --  \\
VIRGOHI27 & S3.6 &      & 0.97 &    --  &  -- &    --  \\ 
VIRGOHI27 & S3.6 & E3.6 & 0.83 &    --  &  -- &    --  \\
\end{tabular}
\caption{Results of the Arecibo follow-up observations of the HIJASS objects without optical counterparts. [1] Object, [2] offsets from the object's HIJASS position in R.A. and Dec.,[3] rms noise level, [4] line flux, [5] profile FWHM, [6] centre velocity. Numbers in square brackets are for detection at less than 4$\sigma$.}
\end{table*} 

\subsection{Objects with no optical counterparts - results from Arecibo}
There have been other surveys of limited areas of sky to relatively large mass and column density limits that have also essentialy found no isolated \HI\ clouds that cannot be associated with optical sources. Thus our detection of four sources is potentially very interesting and a new observation. Previous surveys have covered many different galaxy environments from clusters (Weinberg et al. 1991), groups (Kraan-Kortweg et al. 1999; Dickey 1997) the general field (Spitzak \& Schneider 1998; Henning 1995; Sorar 1994) and voids (Hoffman et al. 1992). The \HI\ clouds in the Hercules Cluster reported by Dickey (1997) were not confirmed by van Driel et al. (2003).  Two possible inter-galactic clouds were thought to have been identified in the past, but both of these are now known to have associated optical galaxies (Giovanelli \& Haynes 1989; Schneider et al. 1983). Kilborn et al. (2000) identified a small \HI\ cloud, but it is now thought to be a local HVC.

Below we describe each of our four sources in more detail. For VIRGOHI4 and VIRGOHI13 the position given is from the Jodrell Bank observations. For VIRGOH21 and  VIRGOH27 the positon is that of the Arecibo aperture that contains the highest flux. \\
{\bf VIRGOHI4} (RA=12h21m26s, Dec=14d24'42'', J2000) - this is a strong signal at about 2200 km s$^{-1}$. The line of sight lies through M86 ($v$=-244 km s$^{-1}$) and so no optical image can be seen. VCC0335 is 2 arc min away, it is listed in LEDA as an SO galaxy with $M_{B}=-13.2$. The measured \HI\ mass is far higher than might be expected of a dwarf SO galaxy (Conselice et al. 2003) so, we conclude that this detection is due to a galaxy that lies behind M86. We observed only the central position at Arecibo, due to time constraints and the relative proximity of the 220 Jy continuum source M87, which degraded the quality of the spectrum. The spectrum is shown in fig 4.\\
{\bf VIRGOHI13} (RA=12h17m51s, Dec=14d46'31'', J2000) - this is a relatively strong signal at 1274 km s$^{-1}$ in the HIJASS data. Surprisingly, this detection was not confirmed at any of the 16 positions we observed at Arecibo, covering a little more than the Jodrell Bank HPBW (fig. 4), with an average 0.95 mJy rms noise level. The $v$=1035 \kms\ \HI\ line signal picked up at the position 9$'$ North of VIRGOHI13, well outside the 6$'$ HPBW radius of the Jodrell Bank telescope, is due to the gas-rich galaxies NGC 4298 and NGC 4302, located \am{11}{8} North of the VIRGOHI13 position. The object would have to be quite extended to avoid detection at Arecibo: with a total HIJASS flux of 2.4 Jy \kms\ over $W_{20}$=100 \kms, the average flux in the line is about 24 mJy, or 25 times the Arecibo rms noise, whereas per Arecibo position the estimated 3$\sigma$ detection limit for a 100 \kms\ wide line is 0.3 Jy \kms. Thus, if spread out over at least 8 independent \am{3}{6} HPBW Arecibo beams, the HIJASS source could in principle remain undetected at the 3$\sigma$ level at Arecibo. Given that VIRGOHI13 is also undetected in the HIPASS data we presume that the 'detection' is peculiar to the HIJASS data and may well be due to NGC4298 or NGC4302 being picked up in a Jodrell bank side lobe. \\
{\bf VIRGOHI21} (RA=12h17m51s, Dec=14d46'31'', J2000) - a relatively weak signal in the HIJASS data at 1966 km s$^{-1}$. At Arecibo we observed a grid of 16 positions surrounding the nominal position of VIRGOHI21, bracketing the \HI\ emission from this object. The largest flux is detected in a beam off-set from the HIJASS position by 2.7 arc min to the west (fig. 4). The object is detected with high signal to noise in five separate beams indicating a source larger than the Arecibo beam. The sum of the fluxes measured in each of the five beams is very close to the total measured by HIJASS. Carrying out a search in NED with a radius of 3 arc min about the Arecibo peak flux position produces just one object, a faint radio continuum source.  \\
{\bf VIRGOHI27} (RA=12h26m45s, Dec=19d44'38'', J2000) - this source has quite a strong peak flux and a narrow velocity width in HIJASS, HIPASS and the Arecibo data. The source lies at $\approx 1652$ km s$^{-1}$. At Arecibo we observed 10 positions around the nominal position of VIRGOHI27, bracketing the \HI\ emission from this object, it was detected in three beams. The centre of the \HI\ source lies 3.6 arc min east of the HIJASS position and again the source appears to be larger than the beam. The sum of the fluxes in the three beams is approximately the same as the HIJASS total. Besides the main gaussian component at 1658 \kms, the spectra at the 2 offset positions (\am{3}{6} East, \am{3}{6} North) and (\am{3}{6} East, \am{7}{2} North) show another component at about 1570 \kms, with a peak flux density of 4 mJy. This component does not appear in the HIJASS spectrum, but the signal is relatively weak and the offset positions are around and outside the HPBW radius of the Jodrell Bank telescope. As these positions were observed on different nights, it does not appear likely that these lines are due to low-level radio interference. The Arecibo spectrum at the peak flux position is shown in fig. 4. A search of NED using a 3 arc min radius from the peak flux position produced two faint 2MASS objects. \\

\begin{figure*}
\subfigure{\psfig{figure=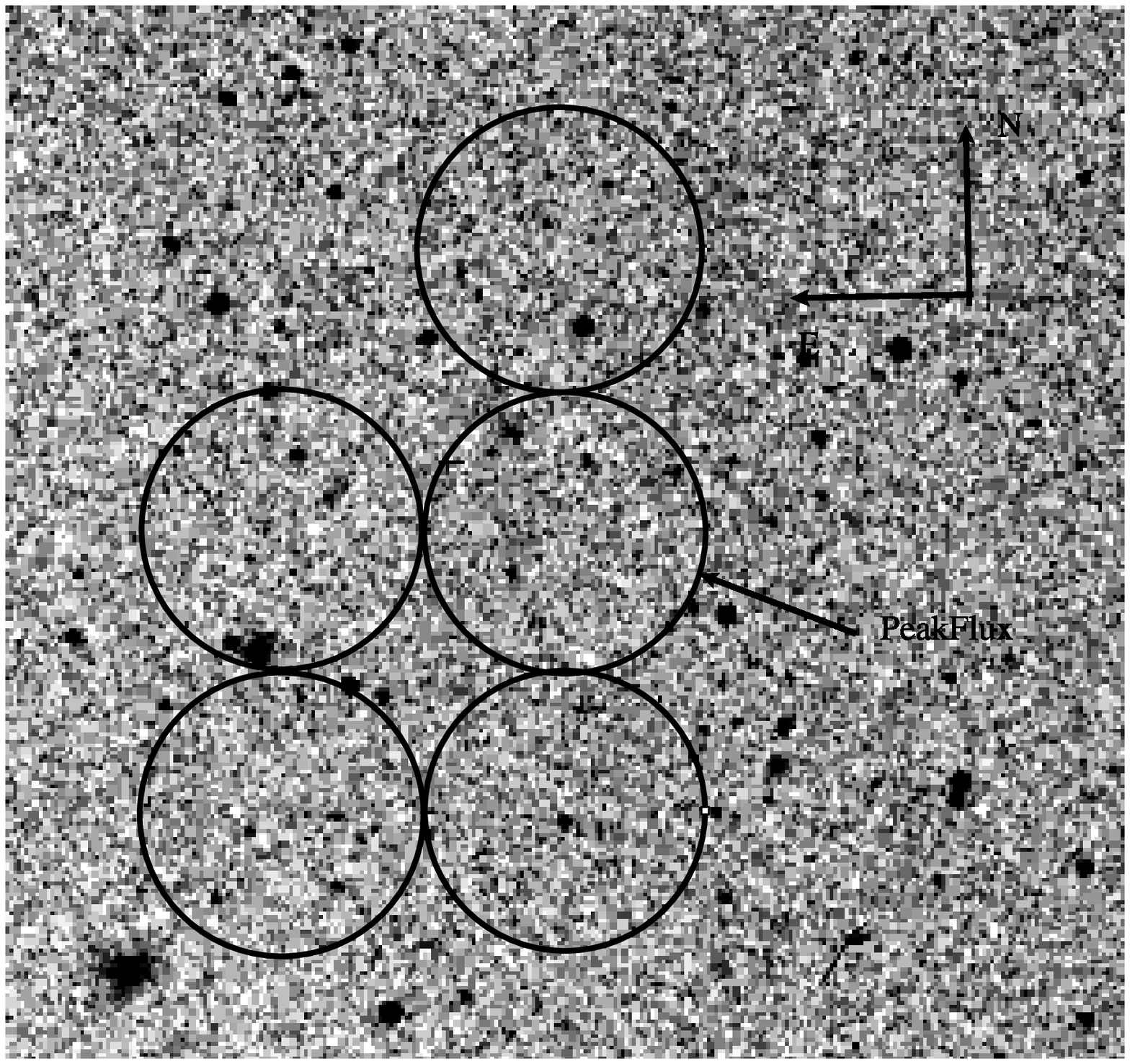,width=8cm}}
\subfigure{\psfig{figure=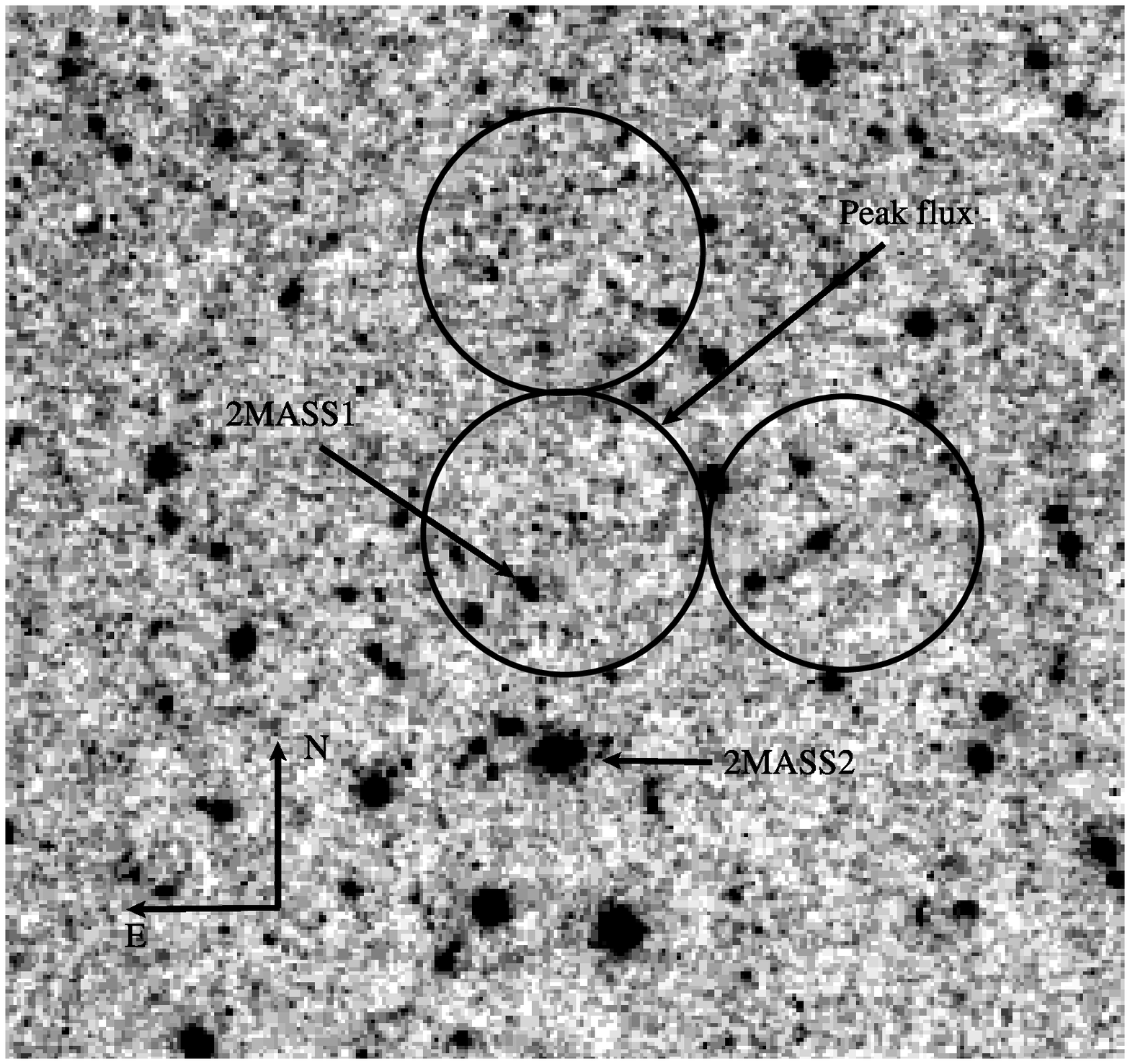,width=8cm}}
\caption
{Digital sky survey images (12'$\times$12') of the fields containing VIRGOHI21 (left) and VIRGOHI27 (right). The fields are centred on the Arecibo peak flux positions. The circles mark the approximate positions of the apertures where a \HI\ detection was made (table 2). The two 2MASS galaxies listed in NED are marked on the image of VIRGOHI27. }
 \end{figure*}

Optical images from the digital sky survey of the area around the two remaining \HI\ detections (VIRGOHI21 and VIRGOHI27) are shown in fig. 5. There does not appear to be any objects that can convincingly be assigned to VIRGOHI21. The emission is extended over an area of about 45 sq arc min which is large compared to the optical sizes of the brightest Virgo cluster galaxies (NED lists a diameter of about 8 arc min for M87). The two 2MASS objects close to the VIRGOHI27 position are labeled in fig. 5. The brighter source (labeled 2MASS2) was not detected in a beam almost centered on its position, so we rule this out as a possible optical counterpart. The optical size of 2MASS1 is very small compared to the HI emission, which extends over about 27 sq arc min. For a (B-K) colour of 2-4 the 2MASS K band magnitude leads to B band magnitudes of -13.7 to -15.7. This would give acceptable values of ($M_{HI}/L_{B}) \approx 0.3-2.0$, though the \HI\ is extended over a comparitively large area.

Giovanelli and Haynes (1989) found a very much larger \HI\ cloud on the outskirts of the Virgo cluster which they described as a 'protogalaxy. This cloud extends along it major axis more than 24 arc min and has an \HI\ mass of $\approx 4 \times 10^{9}$ $M_{\odot}$ (it is very much more extended and massive than the two \HI\ clouds described here). Subsequently a dwarf irregular galaxy was found at one of the peaks of the \HI\ distribution (McMahon et al. 1990) and it is now thought that the \HI\ cloud is associated with this galaxy. This dwarf galaxy is very prominent on the digital sky survey data (easily seen and much larger than the 2MASS galaxies seen in the field of VIRGOHI27). Whether VIRGOHI21 and VIRGOHI27 are isolated HI clouds is still an open question that will only be settled with deep 21cm interferometry, to map the distribution of \HI\, and deep CCD imaging to look for low surface brightness features.

To summarise six objects were detected in the Jodrell Bank survey that had no obvious optical counterparts. Four of these were later confirmed using deeper pointed observation at Jodrell Bank. These four were susquently re-observed at Arecibo. One VIRGOHI4 was confirmed as a galaxy lying behind M86. VIRGOHI13 was not detected at Arecibo and the Jodrell Bank signal is probably due to emission from a bright galaxy in a side lobe. VIRGOHI21 and VIRGOHI27 were confirmed extended \HI\ sources by the Arecibo observations. There are no confincing optical counterparts on the digital sky survey images

\section{The Virgo cluster HI mass function}
From a sample of 1000 field galaxies from the HIPASS survey Zwaan et al. (2003) have recently derived values of $\log{M_{HI}^{*}}=9.8$ and $\alpha=-1.3$ from a Schechter fit to a 'field' galaxy \HI\ mass function. Assuming a distance of 16 Mpc for all of our galaxies we can derive a mass function for galaxies in the cluster environment.  In fig. 6. we compare the shape of these two functions. It is clear that there is a relative shortage of low \HI\ mass galaxies in the Virgo cluster compared to the field (see also Rosenberg and Schneider, 2002). There is a small caveat to this; there is probable incompleteness in the last mass bin, and possible incompleteness in the last but one mass bin, of the mass function. But, we would have to have missed an order of magnitude more galaxies than we have found to make the Virgo and field \HI\ mass functions agree and, we do not believe we have done this. What makes this result very interesting is that the luminosity function of Virgo compared to the field is also very different, but in the opposite sense. Recent determinations of the field galaxy luminosity function give a faint-end slope of $\approx -1.2$ (Norberg et al. 2002; Blanton et al. 2001) and it does not appear to steepen at fainter magnitudes (Roberts et al., (2003). For Virgo we have recently found a slope of -1.7 (Sabatini et al. 2003). Assuming that the cluster and field initially had the same baryonic mass function, it seems that the conversion of atomic gas into stars in the cluster has been very much more efficient than in the field for low mass objects. 

This suggests that gas loss by dwarf galaxies in the cluster environment is less important than in the field. Thus gas removal mechanisms, like ram pressure stripping, which potentially operate on cluster galaxies only (particularly the lowest mass galaxies) cannot be important (see also Sabatini et al., 2003b). We can investigate this further by considering the values of $(M_{HI}/L_{B})$. As stated above it has been known for some time that Virgo cluster galaxies are \HI\ poor compared to field galaxies. Typically, galaxies selected from  \HI\ surveys have tended to have, on average, quite high values of $(M_{HI}/L_{B})$. For example Kilborn et al. (2003) found, for a HIPASS select sample, that $(M_{HI}/L_{B})=1.8-3.2$ for early to late type galaxies. 
This compares to values of 0.1-0.7 for early to late type galaxies selected by their optical properties (Roberts and Haynes (1994).
The mean of our Virgo sample is 0.5 (with large scatter) and there are many galaxies with very small values of $(M_{HI}/L_{B})$ (see table 1). For the Fornax cluster Waugh et al. (2002) find a mean value of $(M_{HI}/L_{B})=1.2$ a factor of two higher than our value for Virgo. In fig 7 we show the relationship between absolute B magnitude and $(M_{HI}/L_{B})$. Although there is a large scatter, there is a clear indication that the lower luminosity galaxies are relatively gas rich compared to the brighter galaxies. A linear fit to our data gives $(M_{HI}/L_{B}) \propto L_{B}^{-1.25}$. This is somewhat steeper than that previously found by us (Davies et al. 2001) or that found by Staveley-Smith et al. (1992) (exponents of -0.4 and -0.3 respectively). This is probably due to the extremely small values of $(M_{HI}/L_{B})$ that some of the brighter galaxies have. Values this low are seldom seen in the \HI\ selected field galaxy population. Although Virgo contains relatively few low \HI\ mass objects compared to the field, those that it does have are the most gas rich cluster galaxies. Although this sample does not include 'undetectable' very gas poor galaxies, it does show that the least likely to be ram pressure strip massive galaxies are the most gas poor of any we detect. We do not believe that this is consistent with ram pressure stripping being a primary gas stripping mechanism. Vallari and Jog (1991) came to a similar conclusion. 

To a mass and column density limit of $5 \times 10^{7}$ $\frac{\Delta v}{50 km s^{-1}}$ $M_{\odot}$ and $3 \times 10^{18}$ $\frac{\Delta v}{50 km s^{-1}}$ atoms cm$^{-2}$ respectively almost all \HI\ in the Virgo cluster is associated with bright optically prominent galaxies. The 2 objects with no obvious optical counterparts amount to only 2\% of the total \HI\ detected in the survey. We can estimate the cluster \HI\ mass density of the cluster by assuming the width of the area sampled is the same as the depth of the cluster to get a volume. This leads to a \HI\ mass density of $3.4 \times 10^{9}$ $M_{\odot}$ Mpc$^{-3}$. This a factor of 50 higher than the \HI\ mass density recently calculated by Zwaan et al., (2003) for the general field. Sandage et al. (1984) calculate a luminosity density within the central 6 degrees of the cluster as $5.6 \times 10^{11}$ $L_{\odot}^{B}$ Mpc$^{-3}$ adjusting this for the steeper luminosity function of Sabatini et al., (2003) gives $1.4 \times 10^{12}$ $L_{\odot}^{B}$ Mpc$^{-3}$. This is $10^{4}$ times higher than the value obtained by Norberg et al., (2002) for the local field population. The $(M_{HI}/L_{B})$ of the cluster is $\approx 250$ times lower than the field. If the cluster is assembled out of infalling field galaxies then where has this gas gone? Given the cluster luminosity function some fraction of it seems to have been converted into additional stars in cluster galaxies. This again seems incompatable with gas ram pressure stripping being a primary environmental evolutionary process, particularly as this should be most efficient for the less massive galaxies.

 Blitz et al., (1999) and Braun and Burton, (1999) have suggested that some HVC are the detectable component of the large numbers of DM halos predicted by CDM models to populate the Local Group. For this also to be true in the Virgo cluster we would have expected a large number of low mass \HI\ clouds with a relatively steep mass function (low mass slope $<-1.5$). In fact we find the opposite, there are decreasing numbers of low mass \HI\ detections. A similar conclusion has been reached by Zwaan (2001) who surveyed 5 nearby galaxy groups and could find no \HI\ clouds unassociated with optically identified galaxies. The HVC hypothesis is either incorrect or the cluster HVC have efficiently converted their gas into optically luminous dwarf galaxies.

Given the observed differences between the cluster and field \HI\ mass and luminosity functions it is interesting to speculate on explanations. The 'feedback' mechaniism that CDM modellers invoke to surpress dwarf galaxy formation cannot explain these differences as this is designed to inhibit dwarf galaxy star formation in all environments. The same applies to ram pressure stripping, but specifically in the cluster environment. In the cluster either the initial conditions were very different to the field (Nature) or there is some mechanism that swithchs off feedback (Nurture) and positively promores star formation in the smallest cluster dark matter halos (this is discussed in much more detail in Sabatini et al., 2003b).

\begin{figure*}
\centerline{\psfig{figure=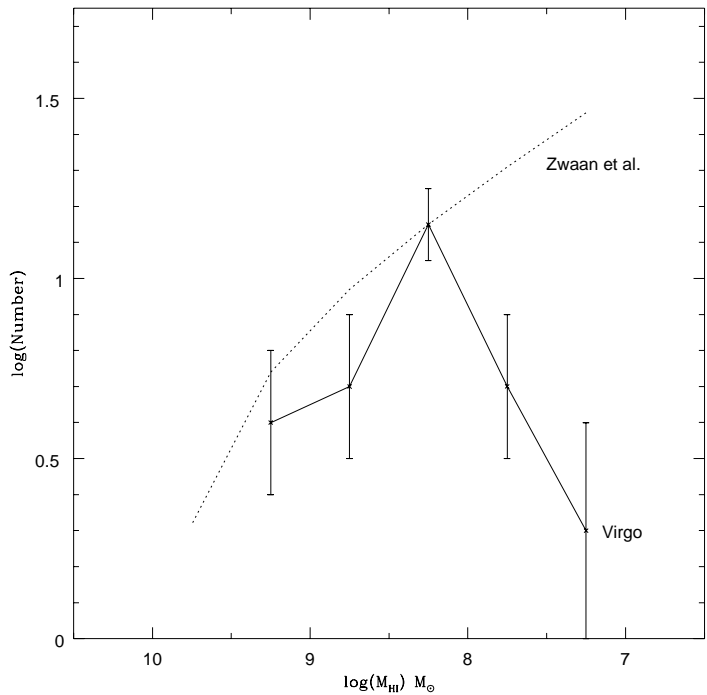,width=10cm}}
\caption
{The \HI\ mass function. The solid line shows the mass function derived from the Virgo data. The dashed line is the derived fit to the field galaxy mass function from Zwaan et al. (2003). The Zwaan et al. data has been arbitarily normalised to the peak in the Virgo data at a mass of $\log{M_{HI}}=8.25$. Note that \HI\ masses may be in error by a factor of four due to uncertainties in distance.}
 \end{figure*}

\section{\HI\ column densities}
An intriguing result of the HIPASS galaxy survey is that \HI\ always seems to be associated with stars - an optical counterpart can always be associated with the \HI\ detection. This is interesting because at first sight it implies that the conditions for star formation are present in a wide range of dark matter halos, from the very large to the very small. An alternative explanation is that it is a selection effect. It has previously been
 demonstrated that star formation thresholds exist such that there is a critical column density at which star formation can proceed (Kennicutt, 1989). These column densities are not very different to the limiting column density of the HIPASS and HIJASS surveys. Thus it is possible that the association of \HI\ and stars is no coincidence, it is what might be expected from a survey that is sensitive to such high column densities. Minchin et al., (2003a) have investigated this idea using a much deeper \HI\ sample. Their much lower column density limited sample ($4.2 \times 10^{18}$ atoms cm$^{-2}$) also failed to find low column density galaxies and all of the detections could be associated with optical counterparts. So it seems that the lack of low \HI\ column density galaxies is real. This result can be explained as an ionisation effect. At about a column density of a few $\times 10^{19}$ atoms cm$^{-2}$ ionisation by the ultra-violet background leads to a dramatic decrease in column density producing a marked 'gap' in column density between those galaxies that are optically thick to the ionising radiation and those that are not (Linder et al., 2003).
The HIPASS and HIJASS surveys and the deep survey of Minchin et al. have focused mainly on the field galaxy population, is the same true for the Virgo cluster ? 

 We compare the \HI\ column densities of our galaxies with those in the field sample of Minchen et al., (2003a). We have calculated the \HI\ column densities of our detections in a similar way (Minchin et al. used $R_{HI}=5R_{e}$ we used $R_{HI}=2.4R_{25}$, these two relations are the same for an exponential disc galaxy with a central surface brightness of 21.7 B$\mu$ (Freeman  1970)). The calculated column densities are listed
in table 1. The minimum detected column density is $\approx 10^{20}$ atoms cm$^{-2}$, almost two orders of magnitude higher than our detection limit. Although many Virgo galaxies are relatively gas poor, compared to their optical brightness they do not have lower column densities than field galaxies. A possible problem with this comparison is that the optical-\HI\ size relation is different for cluster galaxies compared to field galaxies. Intuitively one would think that the cluster galaxies would have tidally truncated radii and so we would have under estimated the column density not overestimated it. Either the same column density limit applies to both field and cluster galaxies or the cluster galaxies have smaller \HI\ sizes and larger \HI\ column densities. In either case we do not find low column density galaxies associated with optically identified galaxies. This conclusion does not apply to the two sources without optical identifications. We can calculate the average column density over the beam width of Arecibo at the peak flux detection points for VIRGOHI21 and VIRGOHI27. In both cases this gives a peak column density of $4 \times 10^{19}$ atoms cm$^{-2}$, about an order of magnitude lower than that of the typical optically identified source (note though that this column density is calculated in a different way, it does not rely on an assumed \HI\ size). It is possible that the two isolated \HI\ clouds have not reached the column density threshold necessary for star formation and that they are examples of the rare objects predicted by Linder et al (2003a) that sit in the gap between the numerous objects that have high (opaque) and low (ionised) measured column densities.

\begin{figure*}
\centerline{\psfig{figure=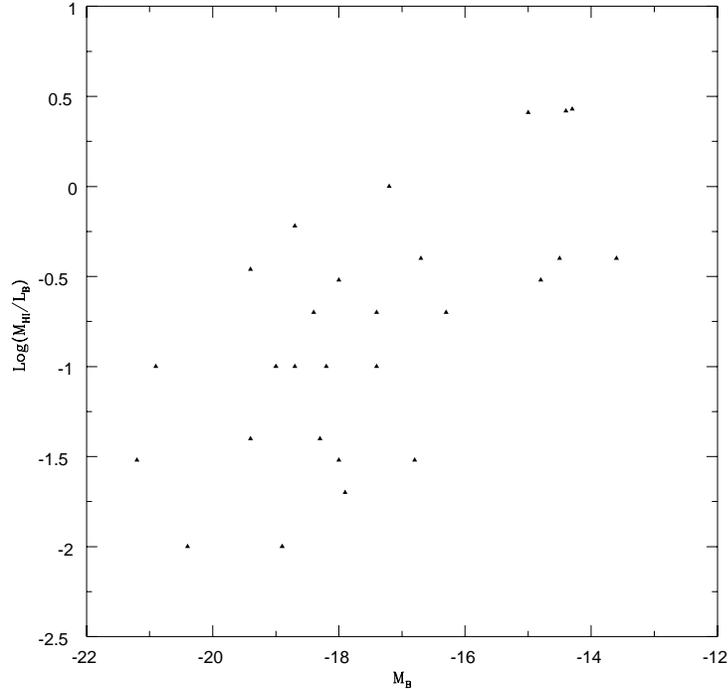,width=10cm}}
\caption
{The absolute magnitude ($M_{HI}/L_{B}$) diagram.}
 \end{figure*}

\section{The velocity distribution}
The distribution of galaxy velocities from our \HI\ sample is shown in fig. 8. It is not the distribution we initially expected. Binggeli et al. (1993) clearly show that the cluster comes to an end at about 2500 $km$ $s^{-1}$, and has a mean velocity of about 1050 $km$ $s^{-1}$. Structurely the Virgo cluster is complex. Binggeli et al. define by their velocity M and W complexes. They also split the cluster into two sub-clusters, A and B. The A cluster falls within our survey area, but its mean velocity is about the same as the cluster as a whole. The \HI\ rich galaxies seem to have the highest velocities and our only explanation is that they are predominately an infalling un-virialised population. A similar conclusion has been reached by Waugh et al. (2002) for the Fornax cluster as they measure a much broader velocity distribution for \HI\ selected galaxies compared to optically selected galaxies (see also Conselice et al. 2003). Conselice et al. (2001) also show that there is a wide spread in both mean velocity and velocity width for Virgo cluster galaxies of different morphological types. The spiral galaxies have a broad velocity distribution with almost constant numbers of galaxies having velocities between 1000 and 2500 km s$^{-1}$, which is consistent with our data. 

\begin{figure*}
\centerline{\psfig{figure=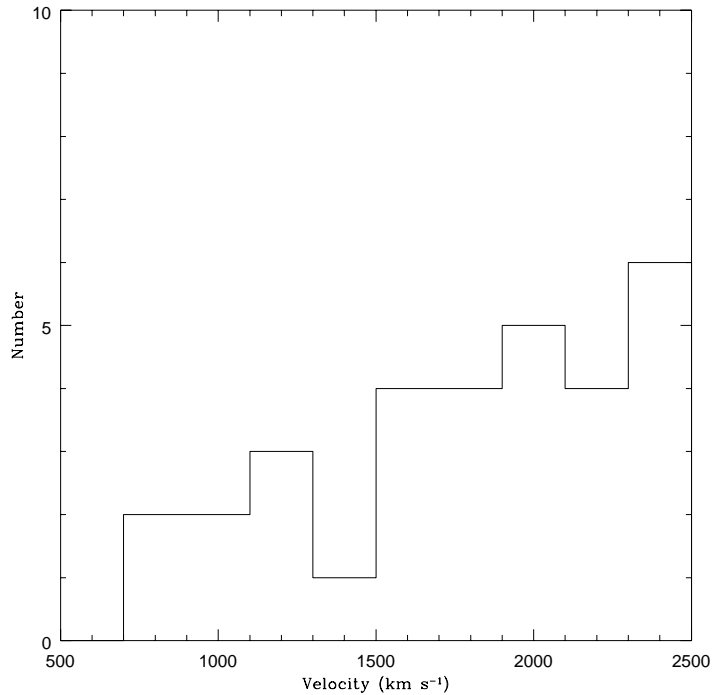,width=10cm}}
\caption
{The distribution of line-of-sight velocities}
 \end{figure*}

\section[]{Conclusions}
\begin{enumerate}
\item To a \HI\ mass limit of $\approx 5 \times 10^{7}$ $\frac{\Delta v}{50 km s^{-1}}$ $M_{\odot}$ and column density limit of $3 \times 10^{18}$ $\frac{\Delta v}{50 km s^{-1}}$ atoms cm$^{-2}$ $\approx98$\% of the \HI\ gas in the Virgo cluster resides in the bright optical sources.
\item There is reasonable evidence to suggest that the cluster \HI\ mass function is very different to that in the field - there are too few low \HI\ mass galaxies.
\item Given the observed steep cluster luminosity function it appears that a larger fraction of \HI\ has been converted into stars in the cluster environment, gas stripping mechanism must be inhibited.
\item The mean \HI\ column density of star forming galaxies is a few times $10^{20}$ atoms cm$^{-2}$ much higher than our calculated column density limit - there are no low \HI\ column density optical sources.
\item Two possible low mass isolated \HI\ objects have been detected in this survey and confirmed with followup observations at Arecibo observatory. They have much lower \HI\ column densities than objects with optical identifications - potentially much lower column densities than those required to form stars.
\item The velocity structure of the Virgo cluster as measured by the gas rich galaxies is very different to that obtained for an optically selected sample - there are proportionally more high velocity objects.
\item The mean value of ($M_{HI}/L_{B}$) for the cluster population is much lower than that for an \HI\ selected field population. The lowest luminosity galaxies are the most gas rich.

\end{enumerate}

\section{acknowledgements}
We want to thank Karen O'Neil and the staff at Arecibo Observatory, especially Tapasi Ghosh, Phil Perillat and Chris Salter, for their help with the observations and data reduction. The Arecibo Observatory is part of the National Astronomy and Ionosphere Center, which is operated by Cornell University under a cooperative agreement with the National Science Foundation. This research has made use of the Lyon-Meudon Extragalactic Database (LEDA), recently incorporated in HyperLeda,  the NASA/IPAC Extragalactic Database (NED) which is operated by the Jet Propulsion Laboratory, California Institute of Technology, under contract with the National Aeronautics and Space Administration

\label{lastpage}

\end{document}